\documentclass{svproc}

\usepackage{url}
\usepackage{amsmath}
\usepackage{graphicx}
\usepackage{subcaption}
\usepackage[export]{adjustbox}
\usepackage{wrapfig}
\usepackage{textgreek}
\usepackage{caption}

\begin{document}
\mainmatter              
\title{Modeling of High and Low Resistant States in Single Defect Atomristors}
\titlerunning{Modeling of High and Low Resistant States in Single Defect Atomristors}  
%
\author{Yuvraj Misra\inst{1} \and Tarun Kumar Agarwal\inst{2}}
\authorrunning{Yuvraj Misra and Tarun Kumar Agarwal.} 
%
\tocauthor{Yuvraj Misra and Tarun Kumar Agarwal}
\institute{Indian Institute of Technology Mandi, Himachal Pradesh, India,\\
\email{B19225@students.iitmandi.ac.in}
\and
Indian Institute of Technology Gandhinagar, Gujarat, India}

\maketitle              

\begin{abstract}
Resistance-change random access memory (RRAM) devices are nanoscale metal-insulator-metal structures that can store information in their resistance states, namely the high resistance (HRS) and low resistance (LRS) states. They are a potential candidate for a universal memory as these non-volatile memory elements can offer fast-switching, long retention and switching cycles, and additionally, are also suitable for direct applications in neuromorphic computing. In this study, we first present a model to analyze different resistance states of RRAM devices or so-called ``atomristors'' that utilize novel 2D materials as the switching materials instead of insulators. The developed model is then used to study the electrical characteristics of a single defect monolayer MoS\textsubscript{2} memristor. The change in the device resistance between the HRS and LRS is associated to the change in the tunneling probability when the vacancy defects in the 2D material are substituted by the metal atoms from the electrodes. The distortion due to defects and substituted metal atom is captured in the 1D potential energy profile by averaging the effect along the transverse direction. This simplification enables us to model single defect  memristors with a less extensive quantum transport model while taking into account the presence of defects.
\end{abstract}
\section{Introduction}
The successful fabrication of a memristor in 2008 at Hewlett Packard Research Labs has opened a window of opportunities towards more energy-efficient computing systems. The term “memristor” is derived from “memory-resistor” and rightly so, as it can work as a non-volatile electronic memory device with multiple memory states \cite{hplabs}. This new class of memory storage unit is a potential candidate for a universal memory that can address the requirements of a future memory device i.e, non-volatile capabilities with fast-switching, long retention and switching cycles \cite{mma}. Metal-insulator-metal (MIM) structures are at the heart of these developing memristive devices, and modelling the electrical characteristics of these nanoscale devices using quantum transport can provide insights into their switching mechanism \cite{cbram}.

The MIM structure consists of electrodes, similar to capacitors, that sandwich an insulating layer, and are usually made of metals like Au, Pt, Al and Ag. The insulating layer in the structure governs many crucial aspects of the device, viz. switching mechanism, conduction mechanism, working voltage polarity, working speed, and resistance ratio \cite{hu}. Recently, 2D sheets of transitional metal dichalcogenides (TMD) such as MoS\textsubscript{2}, MoSe\textsubscript{2}, WS\textsubscript{2}, WSe\textsubscript{2} etc. are being explored to replace transition metal oxides like SiO\textsubscript{2}, HfO\textsubscript{2}, NiO etc, in the switching layer of the future memristive devices \cite{sindef}. It is due to the promise of atomically-thin 2D materials or ``atomristors'' that can mitigate the variability associated with these devices by providing more controlled filament shapes than the bulk oxides \cite{chenvar}. Interestingly, dense memory fabrics, formed by replacing the electrodes in atomristors with graphene, are also crucial to enhancing brain-inspired or neuromorphic computing \cite{mma}.

In atomristors, the switching mechanism has been attributed with the presence of defects in the 2D sheets and metal diffusion through the switching layer, similar to conductive-bridge RRAM (CBRAM) cells \cite{cbram}. It thus becomes imperative to gain an insight into the origin of HRS and LRS in single defect memristors as such an analysis can aid in designing optimum memristive devices through precision defect engineering techniques \cite{sindef}. In this work, we present a model that can capture the transport in the two resistance states with low computational complexity. Firstly, we present the methodology to calculate the electrical characteristics of MIM structures using an in house 1D quantum transport framework, which can solve the 1D Poisson and Schrodinger equation self-consistently within the Non-equilibrium Green's Function (NEGF) framework. Next, the model is validated for the experimentally demonstrated metal - 2D material - metal structures and subsequently used to shed light into the NVRS mechanism. Finally, the model is applied to illustrate the origin of NVRS in atomristors arising due to the presence of a single defect \cite{sindef}. In doing so, we confirm the efficacy of the model by validating obtained results with the experimental data \cite{sindef}.

\section{Methodology}
\begin{figure}[ht]
\centering
\includegraphics[scale=0.30]{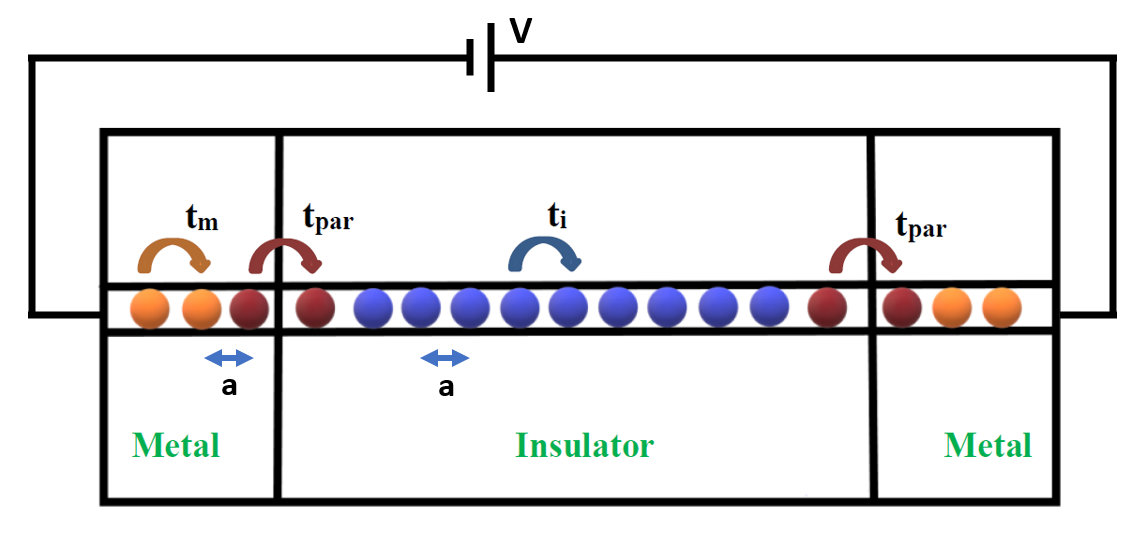}
\caption{Schematic of a MIM structure with finite grid. The arrows indicate the hopping terms in the metal (\textit{t\textsubscript{m}} = 14.03 eV), junction (\textit{t\textsubscript{par}} = 14.73 eV) and insulator (\textit{t\textsubscript{i}} = 15.43 eV) regions. The orange, red and blue points represent the metal , coupled  and insulator  grid points respectively. The inter-atomic spacing \textit{a} equals 0.05 nm.}
\end{figure}

We use the effective-mass tight-binding Hamiltonian approach to model the MIM device, where the eigenstate representation and the real-space representation were used for the transverse and longitudinal dimensions respectively \cite{datta}. Hence, the Hamiltonian matrix was separated into a transverse and a longitudinal part. In working with the NEGF, the matrix of concern was the longitudinal matrix. The first step was, therefore, to write this matrix for our system. This was done using the finite-difference method. The resulting matrix, consisting of the potential energy and hopping term, was then written such that it suitably represented the properties of the device \cite{datta}. This involves accounting for the change in the effective masses of the electrons and hence, the hopping factor in the matrix, according to whether the lattice point lied in the metal or the insulator region. For the points that lied on the metal-insulator boundary, a new hopping term \textit{t\textsubscript{par}} was described to model the metal-insulator interaction (Fig.1). The potential energy for each point was substituted from an estimated potential profile.

The longitudinal Hamiltonian \textit{H} can be expressed as:
\begin{enumerate}
    \item If \textit{$i=j$}
\begin{equation*}
  H_{ij} =
    \begin{cases}
      \textit{$2 t_m$} & \text{$1 \leq i \leq x-2$}\\
      \textit{$2 t_i$} & \text{$x+2 \leq i \leq x+y-2$}\\
      \textit{$2 t_m$} & \text{$x+y+2 \leq i \leq 2x+y$}\\
      \textit{$2 t_{par}$} & \text{otherwise}\\
    \end{cases}       
\end{equation*}
    \item If \textit{$i=j\pm1$}
\begin{equation*}
  H_{ij} =
    \begin{cases}
      \textit{$-t_m$} & \text{$1 \leq \frac{i+j-1}{2} \leq x-1$}\\
      \textit{$-t_i$} & \text{$x+1 \leq \frac{i+j-1}{2} \leq x+y-1$}\\
      \textit{$-t_m$} & \text{$x+y+1 \leq \frac{i+j-1}{2} \leq 2x+y-1$}\\
      \textit{$-t_{par}$} & \text{otherwise}\\
    \end{cases}       
\end{equation*}
    \item Else $H_{ij} = 0$    
\end{enumerate}

Here, \textit{x} and \textit{y} represent the number of metal and insulator grid points; \textit{i} and \textit{j} represent the row and column indices of the Hamiltonian matrix; and \textit{t\textsubscript{m}}, \textit{t\textsubscript{i}} and \textit{t\textsubscript{par}} represent the metal, insulator and metal-insulator junction hopping terms respectively. 

The MIM device was taken to be connected to two contacts. Since the system being studied is an open system, the external effects on the device due to the contacts were considered by adding self-energy matrices, \textit{$\Sigma$\textsubscript{1}(E)} and \textit{$\Sigma$\textsubscript{2}(E)}, to the Hamiltonian for both the contacts. This enables us to solely focus on the device’s working regime \cite{datta}. Next, the local density of states (LDOS) for each lattice point was calculated by computing the spectral function \textit{A(E)}, which was then used to derive the transmission \textit{T(E)}. The quantities were derived using the Green’s function method as shown below \cite{datta}.

\begin{equation}
G(E) = [EI - H - \Sigma\textsubscript{1} - \Sigma\textsubscript{2}]\textsuperscript{-1}
\end{equation}
\begin{equation}
\Gamma\textsubscript{1,2} = i[\Sigma\textsubscript{1,2} - \Sigma\textsubscript{1,2}\textsuperscript{T}]
\end{equation}
\begin{equation}
A(E) = A\textsubscript{1} + A\textsubscript{2} = G\Gamma\textsubscript{1}G\textsuperscript{T} + G\Gamma\textsubscript{2}G\textsuperscript{T} 
\end{equation}
\begin{equation}
T(E) = \textit{trace}(\Gamma\textsubscript{1}A\textsubscript{2}) = \textit{trace}(\Gamma\textsubscript{2}A\textsubscript{1})
\end{equation}

The Poisson equation was coupled with the Schrodinger equation to find the electrical characteristics of the MIM device with the applied bias. The self-consistent solver calculates the density matrix ρ using the NEGF, using an initial estimated potential profile. Finally, a converged electron concentration matrix and potential energy profile are found using the self-consistent solver \cite{newton}.

\section{Application and Analysis}
\subsection{Non-volatile Resistance Switching via Quantum Tunneling Effects}
This section focuses on the study of stable NVRS phenomenon in 2-D transitional metal dichalcogenide (TMD) atomic sheets that arises due to the presence of multiple vacancy defects \cite{nano}. The metal atoms from the electrodes get substituted in these defects and switch the device's resistance from the HRS to the LRS. This switch in resistance can be explained by the increase in tunneling probability in the LRS that occurs due to potential energy distortion in the energy band diagram of the structure. We model this distortion around the defect region by averaging the potential energy profile along the transverse direction. By resorting to such a simplification of the potential profiles, the transport in these devices can be roughly captured using the model.  

In this section, we consider a simplified MoS\textsubscript{2} based memristor (Fig.2) containing multiple defects to illustrate how the simulator can be used to describe its switch from the HRS to LRS using a given potential energy profile. However, the model can be generalized for any material that follows the same HRS and LRS relationships with respect to the applied biases and temperatures. For instance, h-BN, a semi-insulator, which has roughly the same HRS and LRS temperature dependant I-V characteristics as MoS\textsubscript{2} \cite{hbn}.
\begin{figure}[ht]
\centering

\includegraphics[scale=0.30]{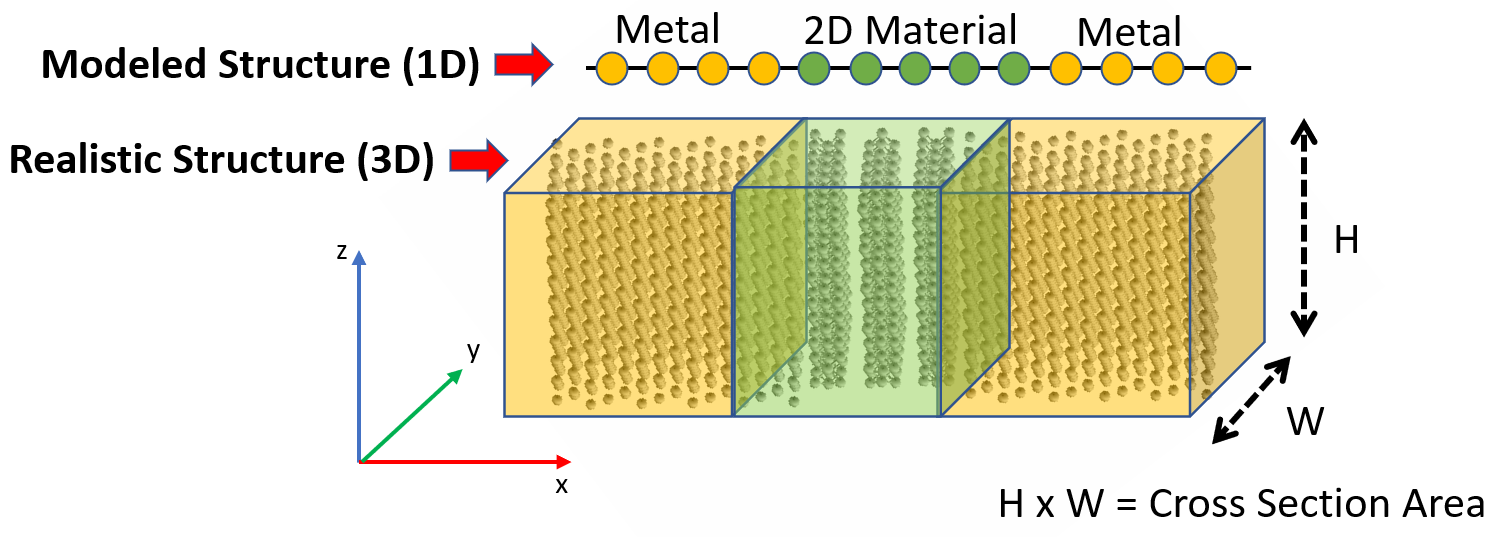} 
\caption{The 1D and 3D visual of a generic 2D materials memristor.}
\end{figure}

\subsubsection{The High Resistance State}
The HRS for MoS\textsubscript{2} can be modelled using a similar potential profile as that of the metal-SiO\textsubscript{2}-metal structure. Given the experimental data for any general MoS\textsubscript{2} memristor with an unknown potential profile, the simulator can be used to determine a rough estimate of the parameters - potential barrier, effective mass of the dielectric, metal-insulator coupling, insulator length, inter-atomic spacing etc. - that allow it to successfully capture the temperature dependent HRS I-V characteristics. Furthermore, it can be used to study the NVRS for a specific 2D materials memristor whose characteristics need to be determined from a known initial potential energy profile. Here, we present a phenomenological analysis of a memristor with multiple defects along two transverse planes to show how the model can capture the device's characteristics.   

In Fig.3, the current characteristics for the HRS are illustrated for different temperature values. The simulator is made to fit the points to a decent extent by tweaking different parameters in the code. Here, a MoS\textsubscript{2} layer is sandwiched between two Au metal electrodes with electron effective mass of 1.1 m\textsubscript{o}. Here, m\textsubscript{o} is the rest electron mass. Based on the fits in Fig.3, an onset potential barrier of 1 eV was extracted for a tunneling effective mass of 1 m\textsubscript{o} and 2D material thickness of 1.5 nm.
\begin{figure}[ht]
\centering

\begin{subfigure}[t]{0.49\linewidth}

\includegraphics[width=0.9\linewidth,height=5cm]{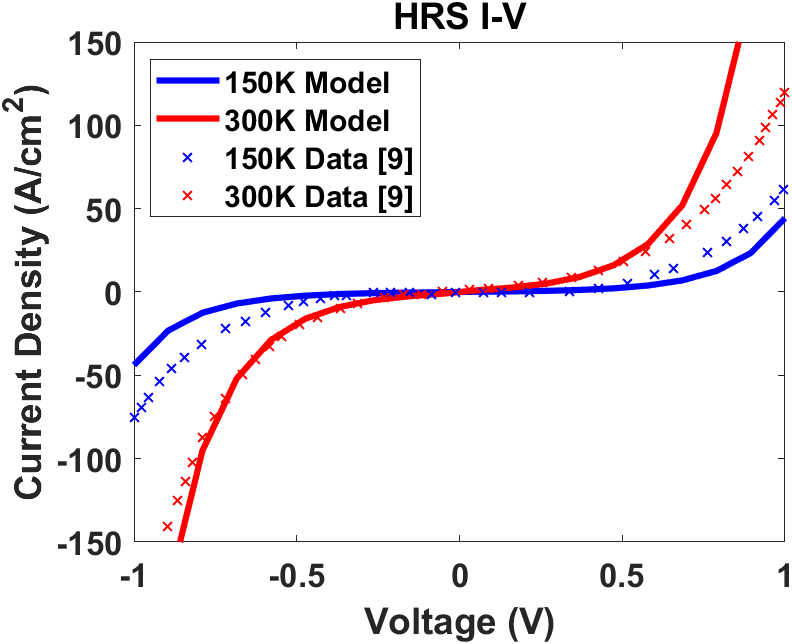} 
\caption{}
\label{fig:subim1}
\end{subfigure}
\begin{subfigure}[t]{0.49\linewidth}
\includegraphics[width=0.9\linewidth,height=5cm]{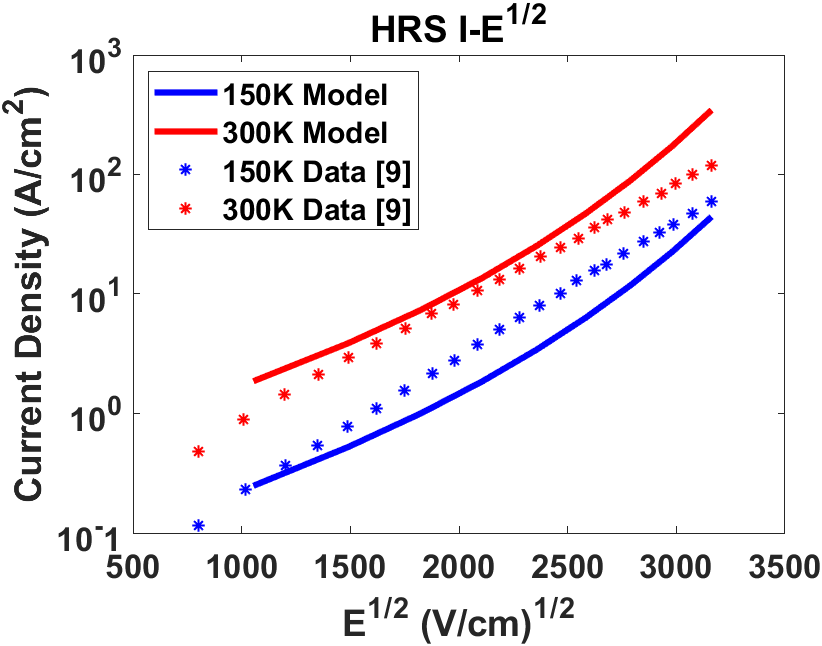}
\caption{}
\label{fig:subim2}
\end{subfigure}
\caption{The High Resistance State I-V characteristics for 150 K and 300 K are shown in (a). The corresponding characteristics as a function of electric field (E) are shown in (b).}
\label{fig:image1}
\end{figure}

For the HRS we observe that the current increases with the temperature. The reason behind this can be explained by recalling the dependency of the Fermi function on temperature. It is known that as the temperature increases, the probability of available density of states to be occupied increases for a given energy. As a result, the current increases. This semiconductor physics also holds true here and hence, the simulator is able to provide reasonable results for the HRS.

\begin{figure}[ht]
\centering

\begin{subfigure}[t]{0.49\linewidth}
\includegraphics[width=0.9\linewidth,height=5cm]{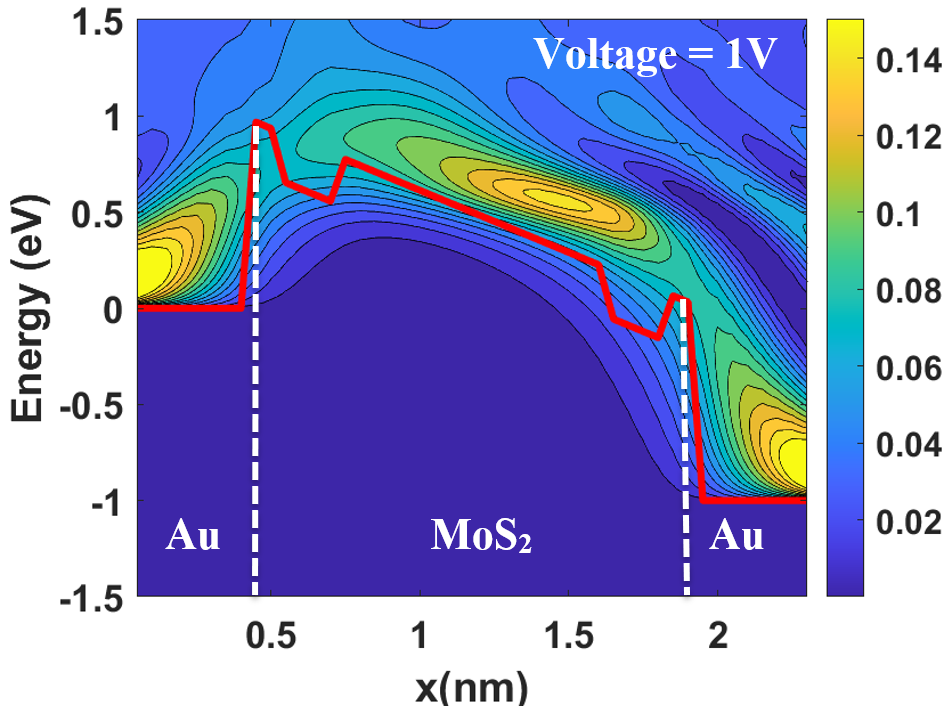}
\caption{}
\end{subfigure}
\begin{subfigure}[t]{0.49\linewidth}
\includegraphics[width=0.9\linewidth,height=5cm]{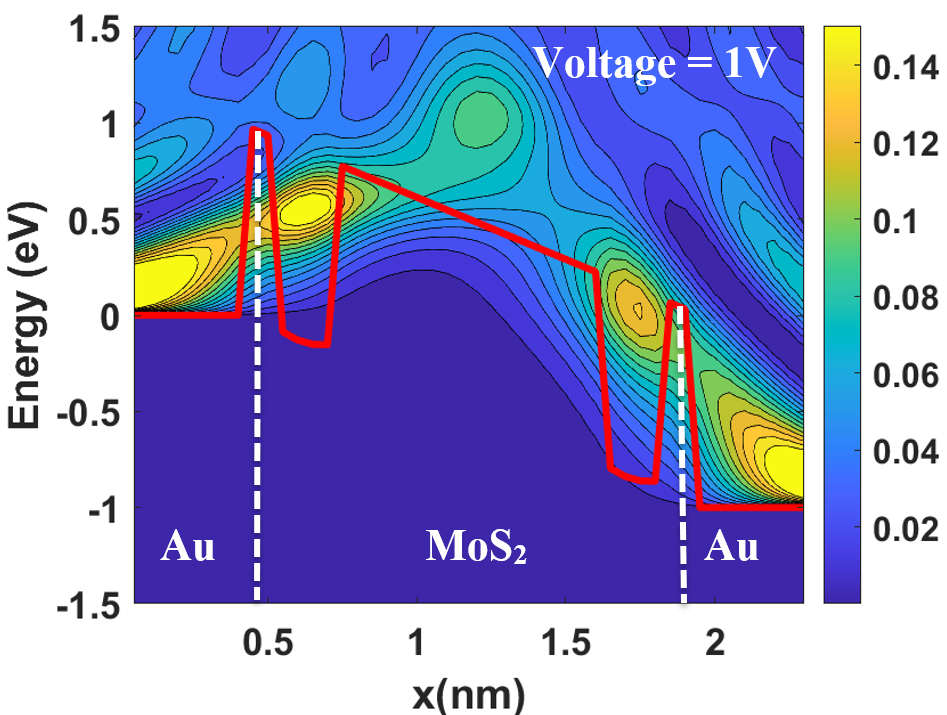} 
\caption{}
\end{subfigure}
\caption{The local density of states for Au-MoS\textsubscript{2}-Au structure (a) in HRS (b) and LRS (after metal substitution) at 1 V. The averaged effect of defects in MoS\textsubscript{2} is highlighted by the potential energy profile shown in red.}
\label{fig:image2}
\end{figure}

\subsubsection{The Switching Mechanism}
The plots in Fig.4 show the switching mechanism from the HRS to the LRS in the simplified MoS\textsubscript{2} memristor due to substitution of sulfur vacancies by gold atoms at 300K. The presence of defects within the MoS\textsubscript{2} bandgap lowers the potential energy as shown in Fig.4-a. When the electric field is sufficiently high, the wells get substituted by metal atoms switching to the LRS. Due to the presence of metal atoms within MoS\textsubscript{2}, the band diagram gets distorted, resulting in deeper wells as shown in Fig.4-b. Beyond a certain extent of depth distortion, energy eigenstates close to the Fermi level are introduced, which allow high transmission through the barrier between the Au-MoS\textsubscript{2} junctions and the wells. This can be seen by the increase in the magnitude of the density of states in the LRS as compared to the HRS in Fig.4. When the wells become deeper they also increase the tunneling probability by reducing barrier width at the Fermi level, which increases the current dramatically.
\subsubsection{The Low Resistance State}
The LRS can be modelled from the simulator by replacing the vacancy-defects in MoS\textsubscript{2} with Au atoms, and subsequently, making the onset at the metal substituted points lower. The distortion of the band diagram due to the coulomb potential relation of MoS\textsubscript{2}, where \textit{r} is the inter atomic distance, has also been captured. As discussed in the last subsection, this leads to the creation of a potential well that allows the electrons at the Fermi level to pass through the allowed energy levels of the potential well, thus improving the tunneling probability. 
For the case under consideration, we replace MoS\textsubscript{2} atoms with Au and set the onset potential to 0 V. The LDOS is then plotted (Fig.4-b) to observe high density state waves form in the wells.  As we increase the doping of the metal in the dielectric at different grid points, we observe more states that the electrons can transmit through. Further, it is observed that the closer the metal substituted defect is to the metal-insulator junction, the higher the density of states is in the potential well. This is because tunneling probability is directly affected by the barrier width and hence, a defect closer to the metal-insulator interface will witness a smaller barrier to transmit through. As a result, the current would increase.  

 It is, however, important to mention that the LRS model cannot capture the temperature dependence of the LRS I-V characteristics as the model only considers the ballistic transport. 
 
 \subsubsection{The HRS and LRS IV Characteristics}
 After getting an insight into the switching mechanism in the previous section, it is now imperative to take a more quantitative approach in explaining the resistance states.   
\begin{figure}[ht]
\centering

\begin{subfigure}[t]{0.49\linewidth}
\includegraphics[width=0.9\linewidth,height=5cm]{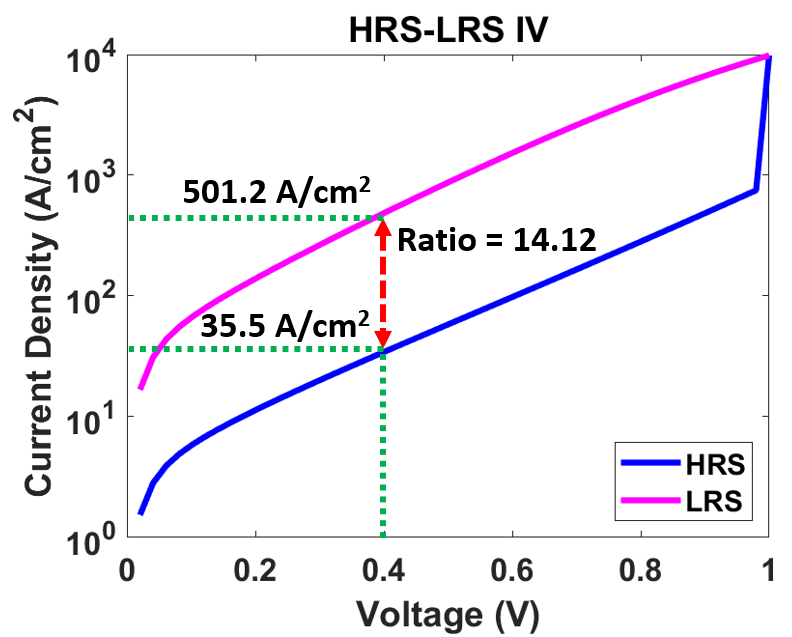} 
\caption{}
\label{fig:subim8}
\end{subfigure}
\begin{subfigure}[t]{0.49\linewidth}
\includegraphics[width=0.9\linewidth,height=5cm]{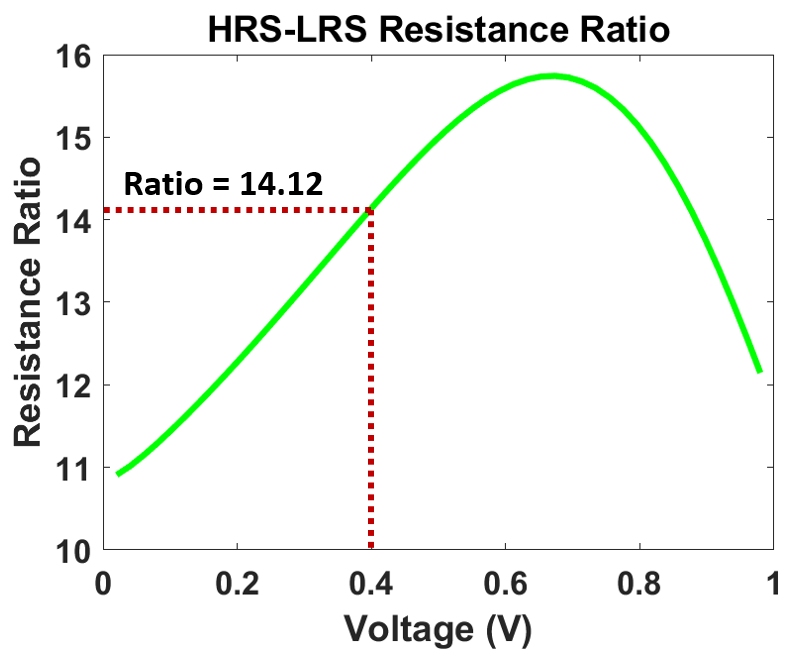}
\caption{}
\label{fig:subim9}
\end{subfigure}
\caption{(a) The electrical characteristics (current density vs. voltage) of HRS and LRS. (b) HRS (R\textsubscript{off}) to LRS (R\textsubscript{on}) resistance ratio for different applied biases}
\label{fig:image3}
\end{figure}
In Fig.5-a, the current-voltage characteristics are plotted for the HRS and the LRS. Firstly, the device is assumed to be operating in the HRS. This is represented by the blue line. As the applied bias is increased, the current increases gradually until the assumed SET voltage of 1 V, where the current increases dramatically due to a change in the resistance state which is modeled as mentioned in the previous section. If the applied bias is decreased from this stage, it follows the magenta curve in the figure, which represents the device's operation in the LRS. Furthermore, Fig.5-b shows the resistance ratio for different applied biases in the device. The expected increase in the ratio with applied bias is observed. The ratio is contained in the range 11 - 15.75.

\section{Transport in Single Defect Memristors}
After gaining an insight on the NVRS of multi-defect 2D materials memristors in the previous section, we now analyze a more specific case - the NVRS of a single defect memristor. Here, the current is passed directly through a defect resulting from a sulfur (S) monovacancy in MoS\textsubscript{2}. The model captures the single defect MoS\textsubscript{2} layer by assuming a 1 nm thick insulating layer - 0.35 nm van der Waals distance on either side of a 0.32 nm thick MoS\textsubscript{2} layer. All other parameters are the same as calibrated in the previous section from the multi-defect memristor data. In this section, we first study the affect of changes in well parameters such as depth and location on the NVRS ratio. The well depth is calibrated in the HRS case as it cannot be ascertained directly from the data; and the well location is varied to capture the rearrangement of atoms that occurs due to metal atom substitution in the defects. Second, we show the single defect NVRS ratio obtained from our model and compare it with the experimental one. 
\begin{figure}[!ht]
\centering
\begin{subfigure}[t]{0.49\linewidth}
\includegraphics[width=0.9\linewidth,height=5cm]{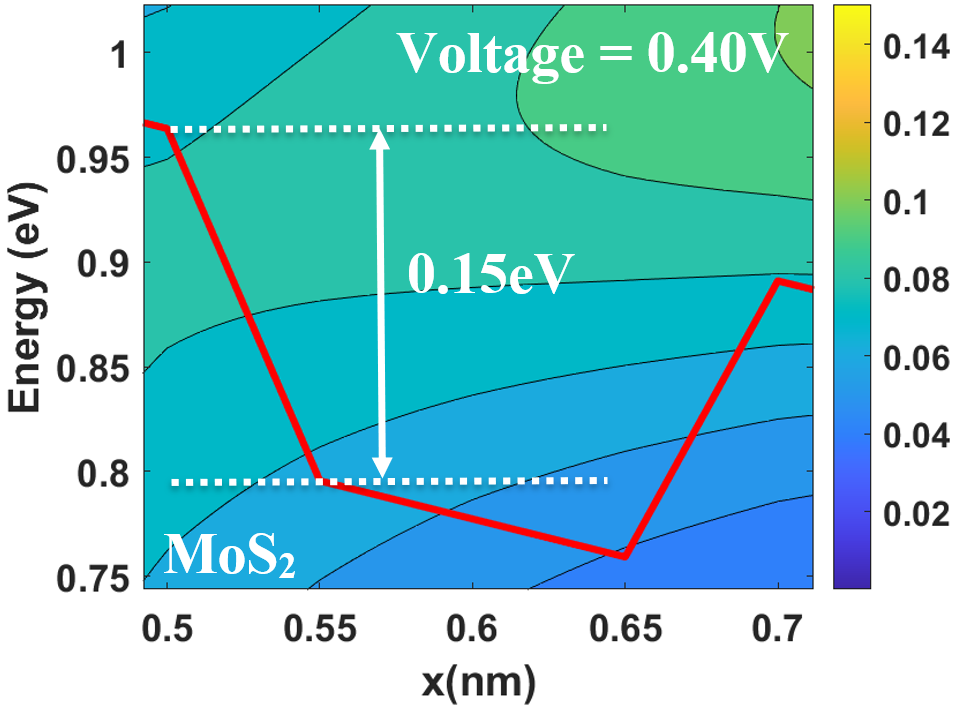} 
\subcaption{}
\label{fig:subim10}
\end{subfigure}
\begin{subfigure}[t]{0.49\linewidth}
\includegraphics[width=0.9\linewidth,height=5cm]{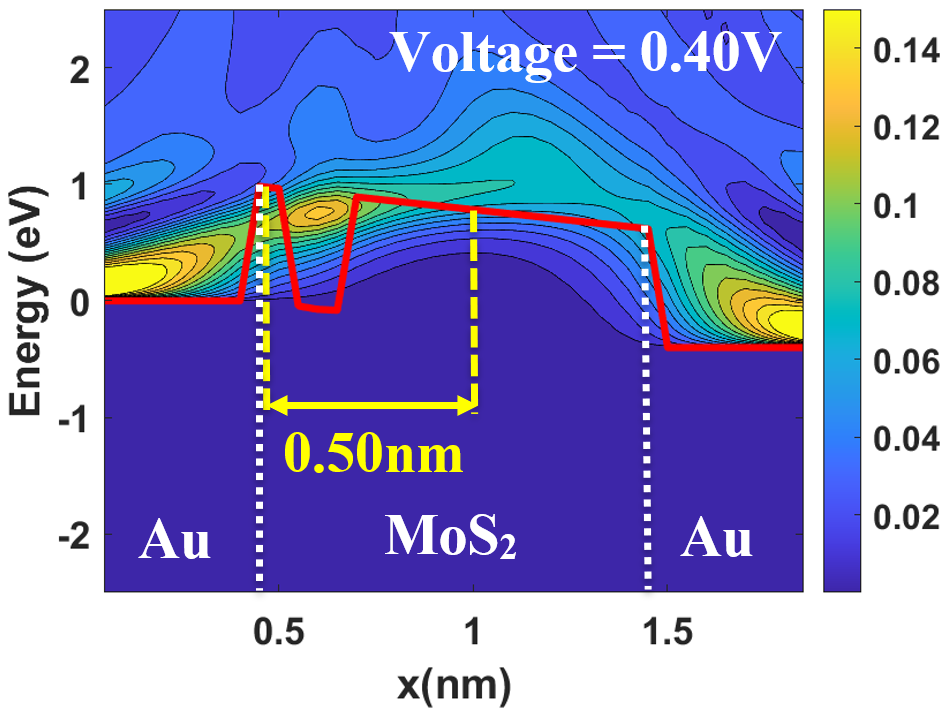}
\subcaption{}
\label{fig:subim11}
\end{subfigure}
\begin{subfigure}[t]{0.49\linewidth}
\includegraphics[width=0.9\linewidth,height=5cm]{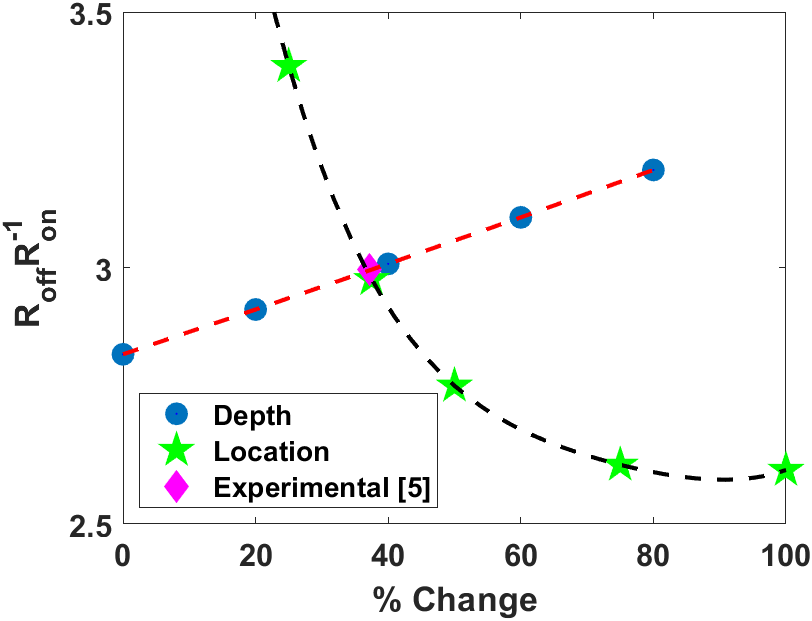}
\subcaption{}
\label{fig:subim12}
\end{subfigure}
\begin{subfigure}[t]{0.49\linewidth}
\includegraphics[width=0.9\linewidth,height=5cm]{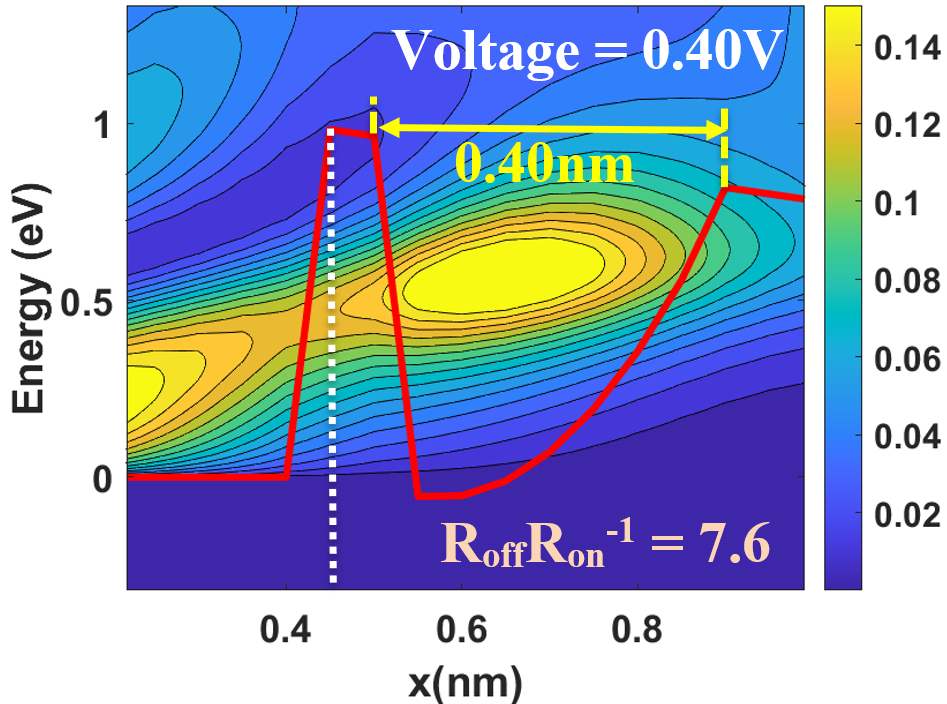}
\subcaption{}
\label{fig:subim13}
\end{subfigure}
\caption{(a) The zoomed in HRS quantum well. Here, the depth is varied in steps of 0.05 eV. (b) The variation in well location within 0.50 nm owing to Mo-S bond length uncertainty. (c) Model calibration by varying well depth and location. The experimental ratio is obtained from Fig.2-f of Ref.\cite{sindef}. The experimental HRS depth is determined to be 0.10 eV. The well location is found to be at 0.18 nm from the left electrode. (d) A possible LRS potential distortion resulting from Au atom substitution.}
\label{fig:image4}
\end{figure}

The well depth parameter corresponds to the depth of the well arising from the monovacancy in the HRS. An initial value of 0.15 eV is chosen for the depth, as shown in Fig.6-a, which is then decremented by steps of 0.05 eV. Each step gives a NVRS ratio as plotted in Fig.6-c. Here, we observe that the trend is almost perfectly linear for the chosen step size of variation and that the ratio ranges between 2.8 - 3.2. Next, we discuss the calibration of the well location. This parameter corresponds to the site in MoS\textsubscript{2} where there is a monovacancy. We observe that the NVRS ratio is sensitive to changes in the monovacancy location (Fig.6-c). In our model, such a variation is considered to be within 0.50 nm i.e. the distance between the left electrode and the Mo atom (Fig.6-b). The change in the ratio with location is roughly exponential and ranges between 2.6 - 3.4. 

After a relationship between the resistance ratio and the parameters is identified, we calibrate our model with the help of experimental observations. We begin by calculating the average switching ratio arising due to Au atom substitution in a S monovacancy from the experimental data given in Fig.2-f of Ref.\cite{sindef}. The extracted experimental resistance ratio comes out to be approximately 3 at 0.40 V. The final calibrated model is then obtained through parameter variation. 

Further, we consider the variation in the LRS potential distortion profile. So far, we had assumed that the sole effect of a gold substitution is the deepening of the wells. However, varying the potential profile after distortion is a parameter that can be experimented with. In our model, we see that distorting the LRS profile as shown in Fig.6-d gives a NVRS ratio of 7.6 indicating that the model is highly sensitive to change in the electrostatic potential profile after the metal substitution. The widening of the well lowers the energy eigenstates. Consequently, the electrons being injected at the Fermi level transmit through the device using these states instead of going over the barrier. Thus, the LRS current increases which results in higher switching ratio. Therefore, the proposed model can provide a physical insight into the HRS and LRS of 2D memristors with low defect densities. 
\section{Conclusion}
A one-dimensional simulator is described to study quantum transport in two-dimensional material memristors by solving the 1D Poisson and Schrodinger equation self-consistently within the Non-equilibrium Green's Function Formalism. It is observed that the model provides a decent fit to the experimental HRS data of multi-defect memristors \cite{nano} and can also capture the temperature dependency of this state. Next, we can use it to study the non-volatile resistance switching ratio (NVRS) of a single defect memristor. For this purpose, a set of two parameters are varied to calibrate the model and gain an insight into the ratio. The efficacy of the calibrated model is proved by capturing the experimental NVRS ratio \cite{sindef} of the single defect memristor. The final model can be used to find the location at which the defect introduces energy eigenstates in the device's potential profile while also providing an insight into the nature of its distortion. It follows from this study that the developed model can be an essential tool in studying the monolayer 2D material memristors with low defect densities especially when alternate simulation techniques are computationally extensive.    
\bibliographystyle{unsrt}
\bibliography{main.bib}

\begin{thebibliography}{10}

\bibitem{hplabs}
J~Joshua Yang, Matthew~D Pickett, Xuema Li, Douglas~AA Ohlberg, Duncan~R
  Stewart, and R~Stanley Williams.
\newblock Memristive switching mechanism for metal/oxide/metal nanodevices.
\newblock {\em Nature Nanotechnology}, 3(7):429--433, 2008.

\bibitem{mma}
Deji Akinwande.
\newblock Memory, {M}emristors, and {A}tomristors.
\newblock {\em IEEE Micro}, 38(5):50--52, 2018.

\bibitem{cbram}
D~Ielmini and V~Milo.
\newblock Physics-based modeling approaches of resistive switching devices for
  memory and in-memory computing applications.
\newblock {\em Journal of Computational Electronics}, 16(4):1121--1143, 2017.

\bibitem{hu}
SG~Hu, SY~Wu, WW~Jia, Q~Yu, LJ~Deng, Yong~Qing Fu, Y~Liu, and Tu~Pei Chen.
\newblock Review of {N}anostructured {R}esistive {S}witching {M}emristor and
  {I}ts {A}pplications.
\newblock {\em Nanoscience and Nanotechnology Letters}, 6(9):729--757, 2014.

\bibitem{sindef}
Saban~M Hus, Ruijing Ge, Po-An Chen, Liangbo Liang, Gavin~E Donnelly, Wonhee
  Ko, Fumin Huang, Meng-Hsueh Chiang, An-Ping Li, and Deji Akinwande.
\newblock Observation of single-defect memristor in an {MoS$_{2}$} atomic
  sheet.
\newblock {\em Nature Nanotechnology}, 16(1):58--62, 2021.

\bibitem{chenvar}
An~Chen and Ming-Ren Lin.
\newblock Variability of resistive switching memories and its impact on
  crossbar array performance.
\newblock In {\em 2011 International Reliability Physics Symposium}, pages
  MY--7. IEEE, 2011.

\bibitem{datta}
Supriyo Datta.
\newblock Nanoscale device modeling: the {G}reen’s function method.
\newblock {\em Superlattices and {M}icrostructures}, 28(4):253--278, 2000.

\bibitem{newton}
Roger Lake, Gerhard Klimeck, R~Chris Bowen, and Dejan Jovanovic.
\newblock Single and multiband modeling of quantum electron transport through
  layered semiconductor devices.
\newblock {\em Journal of Applied Physics}, 81(12):7845--7869, 1997.

\bibitem{nano}
Ruijing Ge, Xiaohan Wu, Myungsoo Kim, Jianping Shi, Sushant Sonde, Li~Tao,
  Yanfeng Zhang, Jack~C Lee, and Deji Akinwande.
\newblock Atomristor: {N}onvolatile {R}esistance {S}witching in {A}tomic
  {S}heets of {T}ransition {M}etal {D}ichalcogenides.
\newblock {\em Nano {L}etters}, 18(1):434--441, 2018.

\bibitem{hbn}
Xiaohan Wu, Ruijing Ge, Po-An Chen, Harry Chou, Zhepeng Zhang, Yanfeng Zhang,
  Sanjay Banerjee, Meng-Hsueh Chiang, Jack~C Lee, and Deji Akinwande.
\newblock Thinnest {N}onvolatile {M}emory {B}ased on {M}onolayer h-{B}{N}.
\newblock {\em Advanced Materials}, 31(15):1806790, 2019.

\end{thebibliography}
\end{document}